\documentclass{iopart}
\usepackage[utf8]{inputenc}
\usepackage{amssymb}
\usepackage{graphicx}
\usepackage[textsize=tiny]{todonotes}

\begin{document}

\title{Enhancing the hybridization of plasmons in graphene with 2D superconductor collective modes}
\author{A. T. Costa$^1$, N. M. R. Peres$^{1,2}$}
\address{$^1$International Iberian Nanotechnology Laboratory, 
4715-330 Braga, Portugal}
\address{$^2$Centro de F\'isica das Universidades do Minho e Porto and Departamento de F\'isica and QuantaLab, Universidade do Minho, Campus de Gualtar, 4710-057 Braga, Portugal
}
\begin{abstract}
We explore ways in which the close proximity between graphene sheets and
monolayers of 2D superconductors can lead to hybridization between their 
collective excitations. We consider heterostructures formed by combinations 
of graphene sheets and 2D superconductor monolayers. The broad range of 
energies in which the graphene plasmon can exist, together with its tunability, 
makes such heterostrucutres promising platforms for probing the many-body physics
of superconductors. We show that the hybridization between the graphene
plasmon and the Bardasis-Schrieffer mode of a 2D superconductor results in
clear signatures on the near-field reflection coefficient of the 
heterostructure, which in principle can be observed in scanning near-field microscopy
experiments.
\end{abstract}
\date{March 2021}

\maketitle

\section{Introduction}

Within the broad class of quantum two-dimensional (2D) materials, superconductors 
are arguably the most challenging, both from the theoretical and the experimental 
perspectives. On the other hand, their complex behaviour is teeming with possibilities, 
from unveiling new physics to providing the basis for disruptive technologies.

As with any other material, elementary excitations can provide insight about the 
fundamental physics of 2D superconductors. The strongly correlated nature of
superconductors endows them with several collective modes, each carrying
complementary pieces of information about the superconducting state. For instance, 
the Higgs mode, associated with oscillations of the amplitude of the superconducting 
order parameter, has energy dispersion~\cite{Littlewood1982,sun2020} 
\begin{equation}
    \hbar\Omega_\mathrm{Higgs}\approx \sqrt{4\Delta^2 + \frac{(\hbar v_F)^2}{d}q^2},
\end{equation}
to leading order in the wave vector $q$. Here, $\Delta$ is the superconducting gap
and $v_F$ is the Fermi velocity and $d$ is the dimensionality of the superconductor. 
Since its energy range lies close to  and above the single particle excitation edge, 
the Higgs mode is damped. Moreover, its coupling 
to far-field electromagnetic radiation is rather weak~\cite{sun2020}, being suppressed
by the small factor $\Delta/E_F$. Back in 1961, Bardasis and Schrieffer proposed~\cite{BardasisSchrieffer1961}
the existence of exciton-like collective modes in superconductors. Their eponymous excitation is
a bound quasiparticle pair, and has a dispersion relation very similar to that of the Higgs mode. 
Its exciton-like character, however, implies its energy lies deep within the superconducting gap, 
making them long-lived excitations. They arise whenever the effective attractive electron-electron 
interaction, responsible for the pairing instability, has competing angular momentum components.
It has been noted in the recent literature that characterization of the Bardasis-Schrieffer mode
can help shed light on the nature of unconventional superconductivity, especially in Fe-based
superconductors~\cite{PhysRevB.103.024519,PhysRevB.92.094506}.

From the recently synthesized 2D superconductors, one of the most promising and intriguing is FeSe, due to the record-high critical temperatures achieved for monolayers~\cite{Wang2012,He2013}. Despite all the activity this system
has attracted, the microscopic mechanisms by which $T_c$ is dramatically enhanced from the modest bulk 
value of $\sim 8$~K to $\sim 65$~K or even $\sim 109$~K~\cite{Ge2015} remain largely unknown~\cite{Huang2017}.

It has been shown recently that combining graphene with superconductors in heterostructures
can lead to fruitful interplay between their collective 
modes~\cite{QNPPerspective,Berkowitz2021,Costa2021}. For instance, hybridization with 
graphene plasmons can enhance the visibility of the superconductor's
collective modes in optical experiments. By carefully designing the geometry of the 
heterostructure, several features of its electromagnetic response can be fine-tuned. 
Moreover, the incorporation of graphene provides a very convenient ``handle'' to modify 
the behavior of the heterostructure during operation, namely its doping level.

\begin{figure}
\includegraphics[width=0.9\columnwidth]{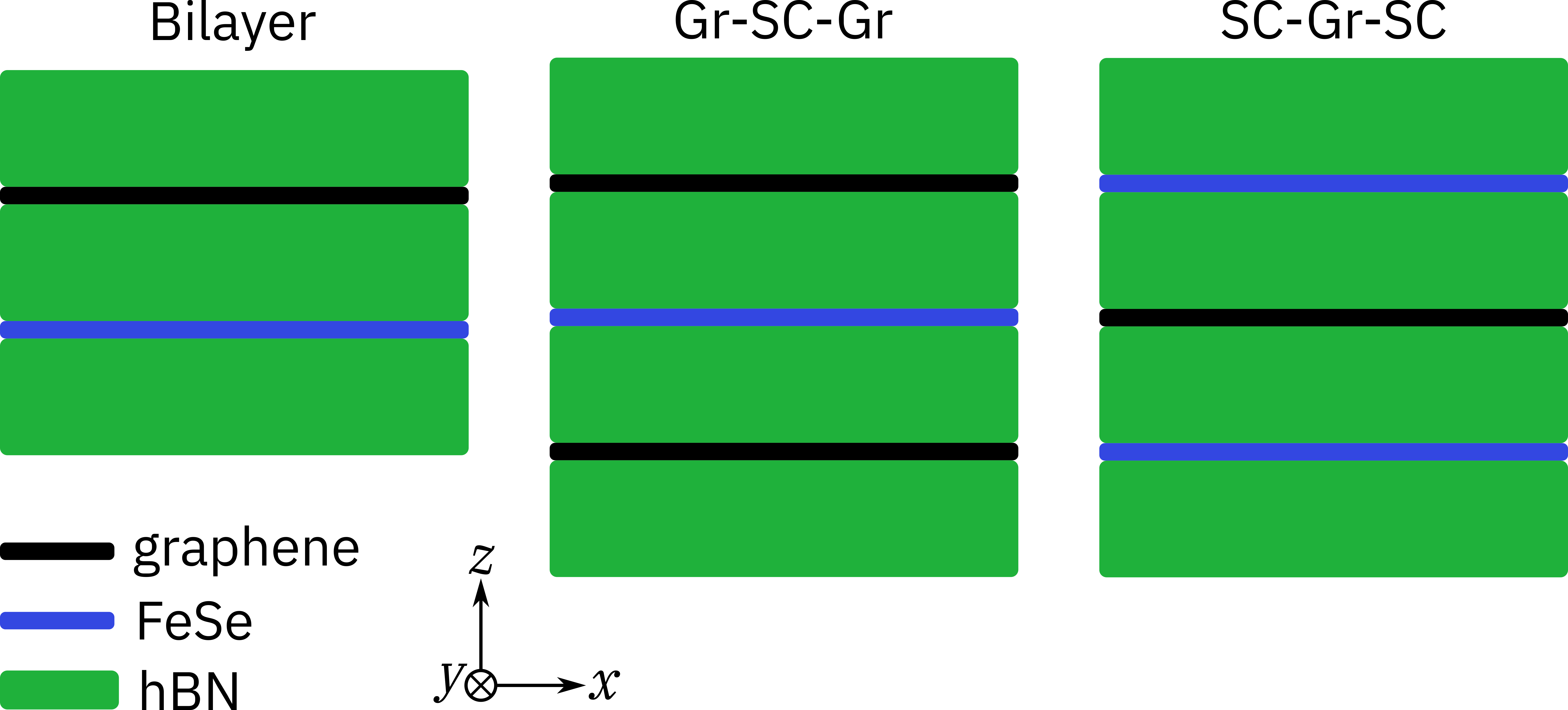}
\caption{Schematic depiction of the three kinds of heterostructures
addressed in this work:}
\label{fig:schem}
\end{figure}

In this paper we study the near-field electromagnetic response of planar
heterostructures combining monolayers of a 2D superconductor and graphene. 
We considered three kinds of heterostructures (shown schematically in 
figure~\ref{fig:schem}): graphene-SC bilayers, graphene-SC-graphene sandwiches 
and SC-graphene-SC sandwiches. In all of them, a uniaxial dielectric 
(such as hexagonal boron nitride) is assumed as spacer between graphene 
and superconductor monolayers or between two superconductor monolayers.
We calculate the heterostructure's reflection
coefficient  associated with the incidence of $p$-polarized waves by 
solving Maxwell's equations subject to appropriate boundary conditions.
The optical properties of hBN are incorporated into the
calculations through its relative permittivity tensor, 
$\epsilon_\mathrm{hBN}=\mathrm{diag}(\epsilon_x,\epsilon_y,\epsilon_z)$.
Given the two-dimensional character of both graphene and the 2DSC, their properties
only enter the calculations through BC. We model both graphene and 
the 2DSC by their non-local optical conductivity tensors~\cite{GoncalvesPeresBook,NunoNonLocalPRB,sun2020}. 
For the superconductor we consider contributions coming 
from the Higgs mode and the Bardasis-Schrieffer mode~\cite{sun2020}. As noted above, 
their features are directly tied to parameters that characterize the superconducting state, 
such as the superconducting gap and its symmetry, which makes them valuable probes into the 
nature of unconventional superconductivity. Moreover, their small dispersion in energy
makes them perfect candidates for hybridizing with graphene plasmons~\cite{Costa2021}. 
The parameters that characterize FeSe were taken from reference~\cite{He2013}. 
There, through a multi-step annealing procedure, it was possible to change the carrier 
density from $n\approx 0.07$ to $n\approx 0.12$ electrons per Fe atom. They observe a 
pure superconducting phase for $n\gtrsim 0.1$ electrons per Fe atom. 

Our results show a strong hybridization between the Bardasis-Schrieffer mode and 
the graphene plasmon, specially in the Gr-SC-Gr geometry. All geometries allow for
tuning the optical response by changing either the heterostructure 
geometry (the thickness of the hBN spacer layers) or graphene's doping level.
The SC-Gr-SC geometry displays long-lived hybrid modes, which can be
relevant for future applications. Moreover, we show that
the hybridized modes impart their signature to the Purcell factor, which
can be probed in scanning near-field optical microscopy (SNOM) experiments~\cite{Ni:2018}.

\section{Bilayers}
\begin{figure}
    \centering
    \includegraphics[width=0.4\columnwidth]{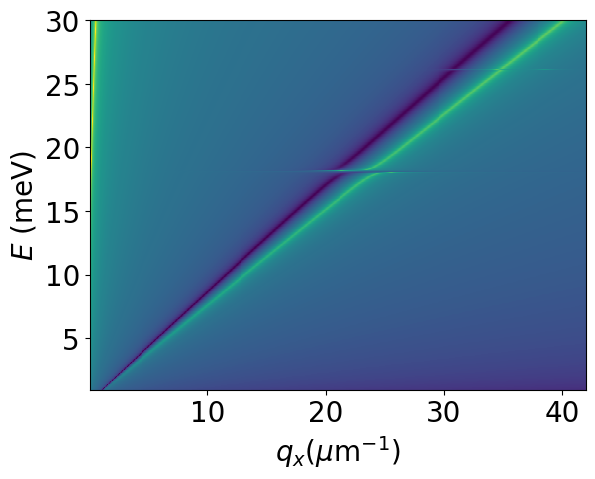}\ 
    \includegraphics[width=0.4\columnwidth]{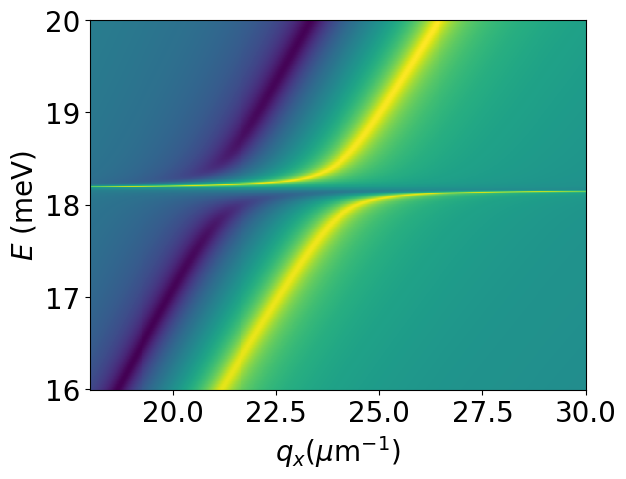}\\
    \includegraphics[width=0.4\columnwidth]{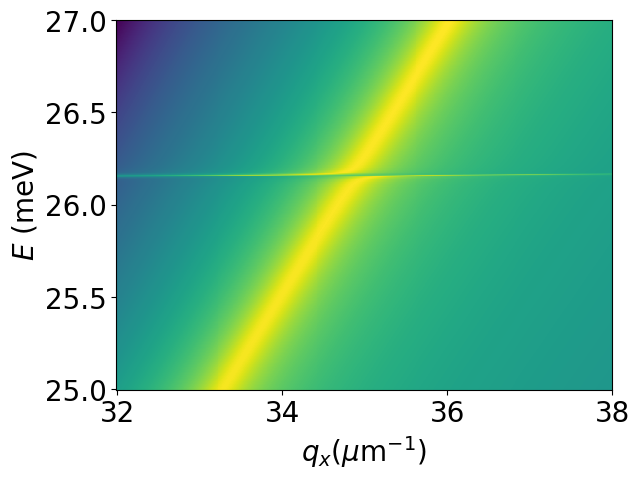}
    \caption{Left: Graphene-SC distance $d=4$~nm, $E_F^{Gr}=400$~meV, $\gamma_\mathrm{Gr}=1$~meV. Superconducting gap $\Delta=13$~meV, $\hbar\omega_\mathrm{BaSc}(0)=18.2$~meV. Right: Zoom at the region of the crossing between the Bardasis-Schrieffer mode and the GPP. Bottom: Zoom at the region of the crossing between the Higgs mode and the GPP.}
    \label{fig:bilayer}
\end{figure}

Here we consider the simplest heterostructure containing one graphene sheet and one
monolayer of FeSe, separated by a thin layer of the insulator hBN of thickness $d$. 
We calculate the frequency and wave vector dependent reflection coefficient to look for signatures of the FeSe collective modes. In fig.~\ref{fig:bilayer} we show the spectrum
of electromagnetic waves as revealed by the imaginary part of the reflection coefficient 
for $p$-polarized waves, $\Im r_p$. By zooming into the spectral region where the
SC collective modes live we can get a clear picture of their hybridization with the
graphene plasmon. It is known that the coupling of the Bardasis-Schrieffer mode
with the electromagnetic field is stronger than the corresponding Higgs' coupling 
by a factor of $E_F/\Delta\gg 1$~\cite{sun2020}. This is reflected by the size of
the splitting between branches around the respective anticrossings, seen in fig.~\ref{fig:bilayer}b and c. In the case of the Higgs mode, the tiny splitting 
between branches ($\sim 85\ \mu$eV) makes it harder to observe it than the 
Bardasis-Schrieffer mode.

\section{Planar cavities}
Here we consider two kinds of planar cavities: i) FeSe sandwiched between two graphene sheets, ii) or a graphene sheet is sandwiched between two monolayers of FeSe. Again, the separation between graphene and FeSe is achieved by a thin layer of hBN.

\subsection{Gr - FeSe - Gr}

\begin{figure}
    \centering
    \includegraphics[width=0.4\columnwidth]{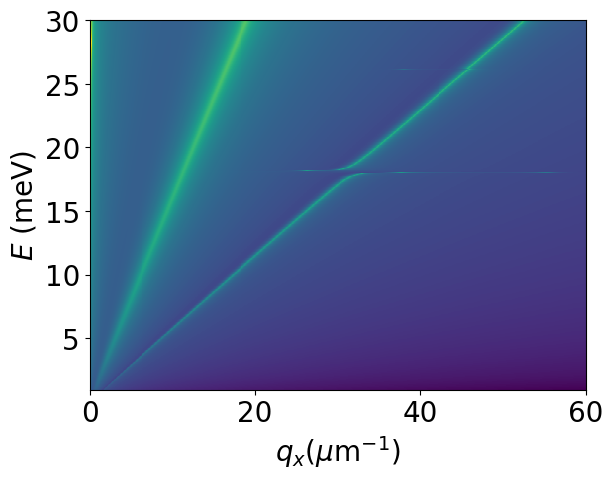}\ 
    \includegraphics[width=0.4\columnwidth]{r_GrSC_EFGr_0.4eV.png}\\
    \includegraphics[width=0.4\columnwidth]{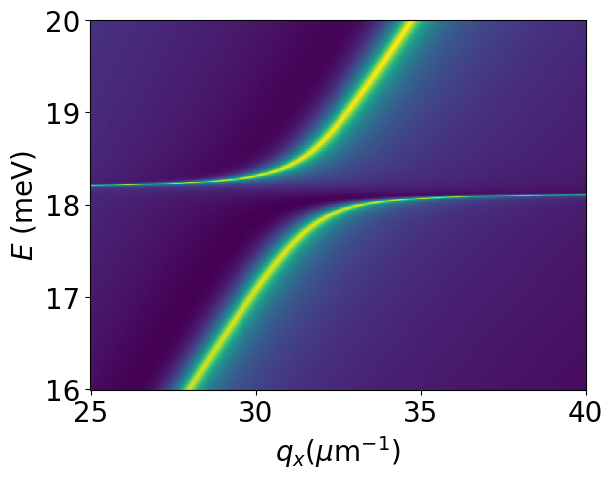}\
    \includegraphics[width=0.4\columnwidth]{r_GrSC_EFGr_0.4eV_zoom_BaSc.png}\\
    \includegraphics[width=0.4\columnwidth]{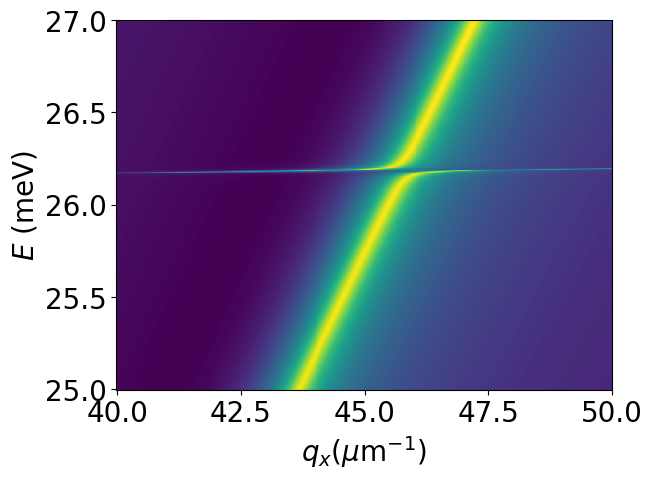}\
    \includegraphics[width=0.4\columnwidth]{r_GrSC_EFGr_0.4eV_zoom_Higgs.png}
    \caption{Top-left: Gr-SC-Gr, top-right: Gr-SC. In both cases Gr-SC distance $d=4$~nm, $E_F^{Gr}=400$~meV, $\gamma_\mathrm{Gr}=1$~meV. Superconducting gap $\Delta=13$~meV, $\hbar\omega_\mathrm{BaSc}(0)=18.2$~meV. Middle: Zoom at the region of the crossing between the Bardasis-Schrieffer mode and the GPP; left: Gr-SC-Gr, right: Gr-SC. Bottom: Zoom at the region of the crossing between the Higgs mode and the GPP; left: Gr-SC-Gr, right: Gr-SC.}
    \label{fig:bi-tri-comparison}
\end{figure}

\begin{figure}
    \centering
    \includegraphics[width=0.4\columnwidth]{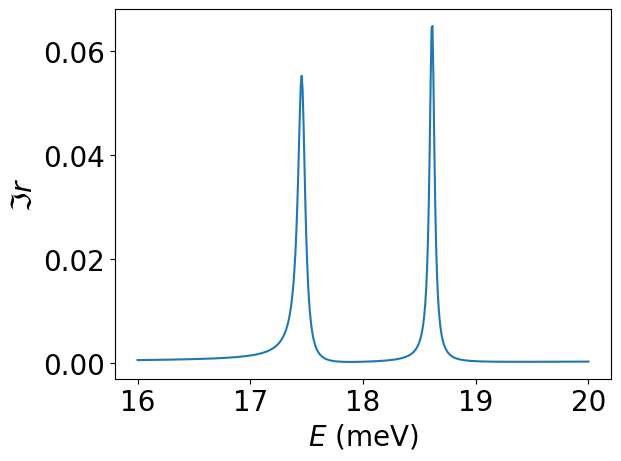}\ 
    \includegraphics[width=0.4\columnwidth]{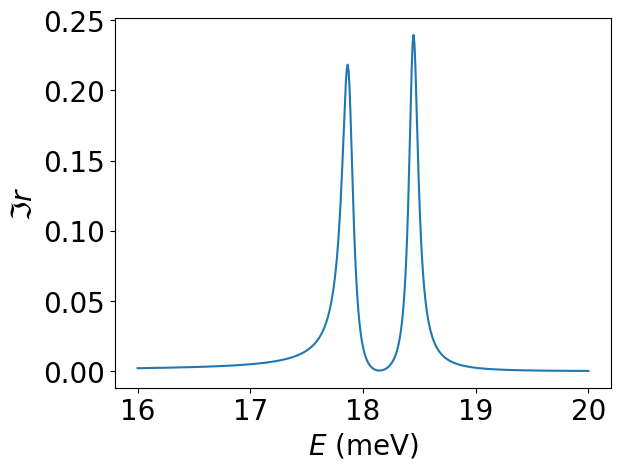}\\
    \includegraphics[width=0.4\columnwidth]{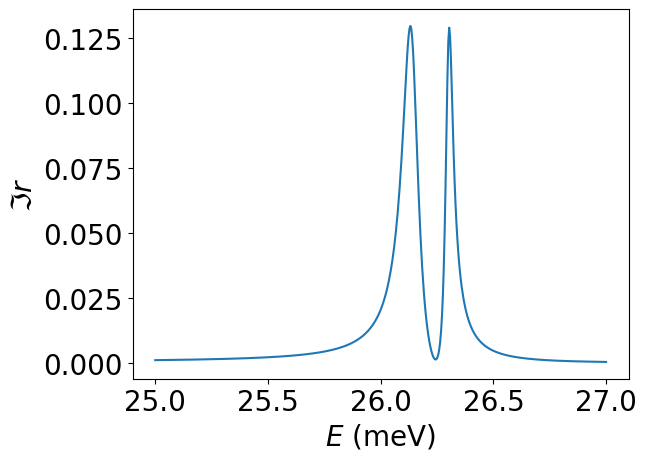}\ 
    \includegraphics[width=0.4\columnwidth]{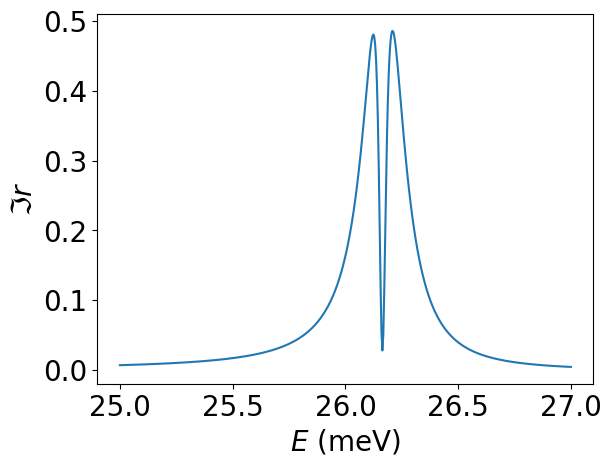}
    \caption{Energy cuts along fixed wave vector, across the Ba-Sh (top) and Higgs mode (bottom) anticrossings. Left: Gr-SC-Gr, $q^\mathrm{BaSc}_\parallel=44\ \mu$m$^{-1}$, $q^\mathrm{Hggs}_\parallel=64.3\ \mu$m$^{-1}$ ; right: Gr-SC, $q_\parallel^\mathrm{BaSc}=24\ \mu$m$^{-1}$, $q^\mathrm{Hggs}_\parallel=34.9\ \mu$m$^{-1}$ . In both cases Gr-SC distance $d=4$~nm, $E_F^{Gr}=400$~meV, $\gamma_\mathrm{Gr}=1$~meV. Superconducting gap $\Delta=13$~meV, $\hbar\omega_\mathrm{BaSc}(0)=18.2$~meV.}
    \label{fig:bi-tri-comparison-wcut}
\end{figure}

\begin{figure}
    \centering
    \includegraphics[width=0.4\columnwidth]{r_GrSCGr_EFGr_0.4eV_zoom_BaSc.png}\
    \includegraphics[width=0.4\columnwidth]{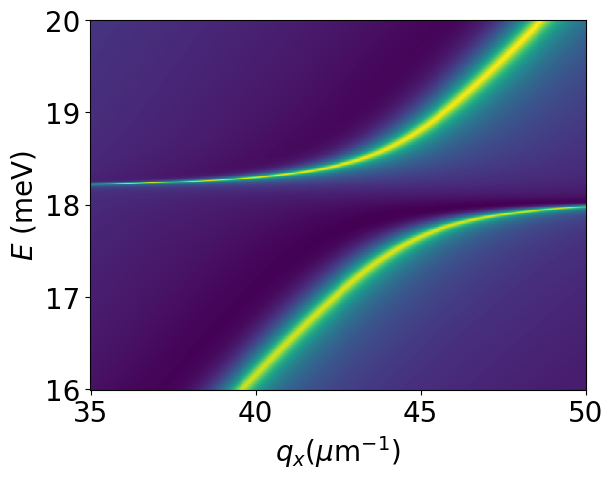}\\
    \includegraphics[width=0.4\columnwidth]{r_GrSCGr_EFGr_0.4eV_zoom_Higgs.png}\
    \includegraphics[width=0.4\columnwidth]{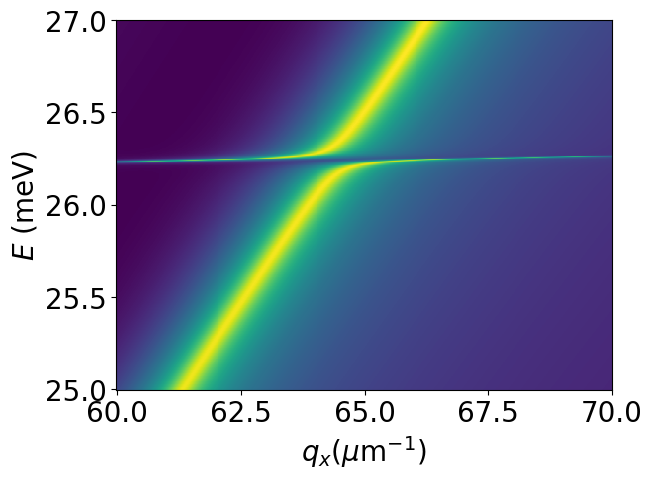}
    \caption{Effect of changing the Gr-SC distance in the sandwich geometry (Gr-SC-Gr). Left: $d=4$~nm, right: $d=2$~nm. Top: Zoom at the region of the crossing between the Bardasis-Schrieffer mode and the GPP. Bottom: Zoom at the region of the crossing between the Higgs mode and the GPP. $E_F^{Gr}=400$~meV, $\gamma_\mathrm{Gr}=1$~meV. Superconducting gap $\Delta=13$~meV, $\hbar\omega_\mathrm{BaSc}(0)=18.2$~meV.}
    \label{fig:changing-distance}
\end{figure}

When compared with the results for the bilayer geometry, we see an increase 
in the splitting between the branches around the anticrossing, both for the Bardasis-Schrieffer 
and for the Higgs mode (fig.~\ref{fig:bi-tri-comparison}). Also noticeable is 
a reduction of the linewidth of both branches (better seen in an energy cut 
along a fixed wave vector, as in fig~\ref{fig:bi-tri-comparison-wcut}. This 
means that the sandwich geometry produces longer-lived hybrid modes than
the simpler bilayer geometry, which could be important for applications.
Also relevant for potential applications is the fact that the features of the
hybrid modes can be controlled electrostatically via graphene's doping level.
Changing the thickness of the insulator layers provides one way to control the features
of the hybrid modes, as seen in fig.~\ref{fig:changing-distance}.

\begin{figure}
    \centering
    \includegraphics[width=0.9\columnwidth]{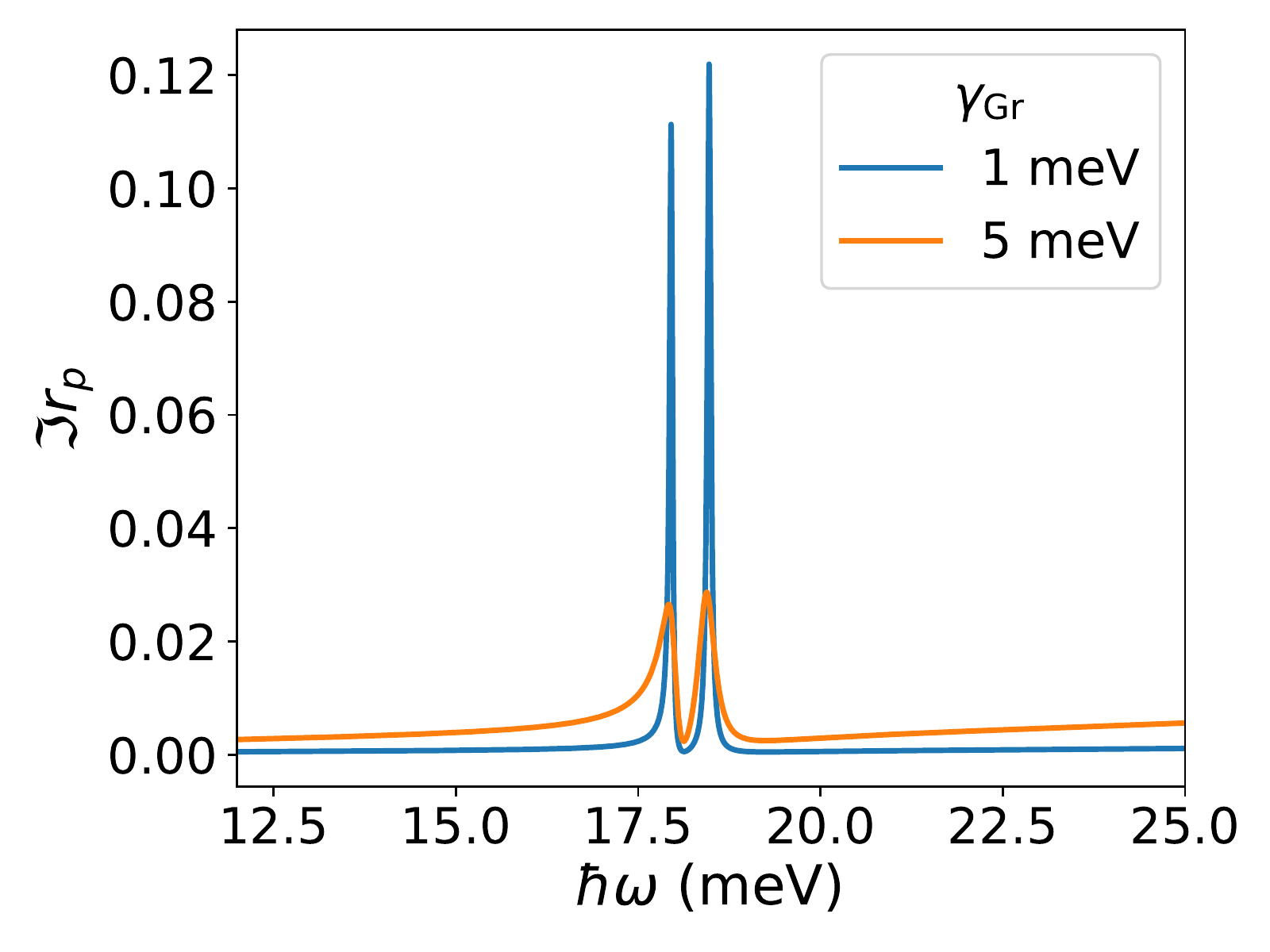}
    \caption{Effect of increasing the graphene relaxation rate $\gamma_\mathrm{Gr}$ on
    the visibility of the hybrid graphene plasmon -- Bardasis-Schrieffer mode. We plot
    the imaginary part of the reflection coefficient as a function of energy $\hbar\omega$
    at a fixed wave vector $q_\parallel=20\ \mu\mathrm{m}^{-1}$. The graphene-SC distance
    is 10~nm and the graphene doping level is $E_F^\mathrm{Gr}=500$~meV.}
    \label{fig:r_x_gamma}
\end{figure}

In order to enhance the visibility of the features associated with the hybridized modes,
we have adopted a fairly small relaxation rate for graphene ($\gamma_\mathrm{Gr}=1$~meV).
Nevertheless, the hybrid modes can still be clearly seen at higher relaxation rates,
as shown in figure~\ref{fig:r_x_gamma}. In fact, those linewidths are considerably smaller than $\gamma_\mathrm{Gr}$. For $\gamma_\mathrm{Gr}=1$~meV the linewidths are $\sim 0.05$~meV,
and for $\gamma_\mathrm{Gr}=5$~meV we find linewidths $\sim 0.2$~meV. This means that
observation of the hybrid graphene plasmon -- Bardasis-Schrieffer mode is possible
with existing graphene preparation techniques.

\begin{figure}
    \centering
    \includegraphics[width=0.4\columnwidth]{r_GrSCGr_EFGr_0.4eV_zoom_BaSc.png}\
    \includegraphics[width=0.4\columnwidth]{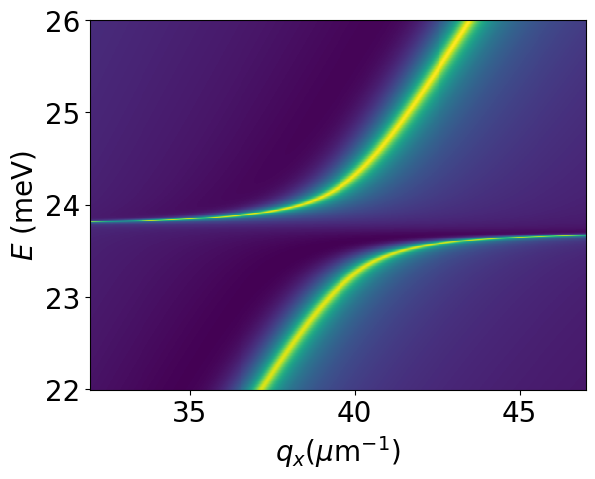}\\
    \includegraphics[width=0.4\columnwidth]{r_GrSCGr_EFGr_0.4eV_zoom_Higgs.png}\
    \includegraphics[width=0.4\columnwidth]{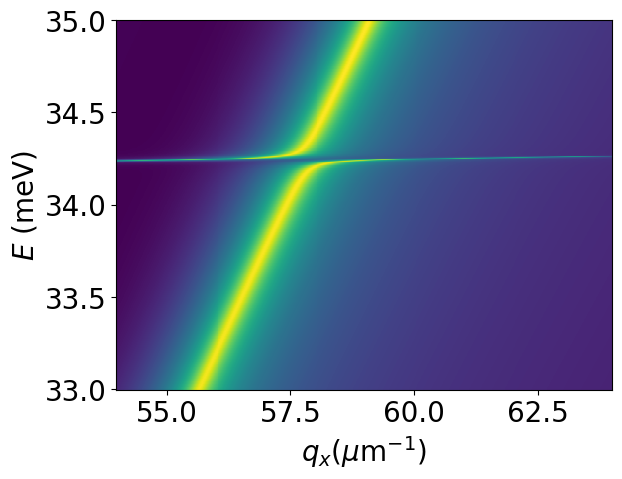}
    \caption{Effect of changing the SC carrier density (Gr-SC-Gr). Left: $n=0.74$~nm$^{-2}$, $\Delta=13$~meV, $\hbar\omega_\mathrm{BaSc}(0)=18.2$~meV. Right: $n=0.81$~nm$^{-2}$, $\Delta=17$~meV, $\hbar\omega_\mathrm{BaSc}(0)=23.8$~meV. Top: Zoom at the region of the crossing between the Bardasis-Schrieffer mode and the GPP. Bottom: Zoom at the region of the crossing between the Higgs mode and the GPP. $E_F^{Gr}=400$~meV, $\gamma_\mathrm{Gr}=1$~meV. }
    \label{fig:my_label}
\end{figure}

\begin{figure}
    \centering
    \includegraphics[width=0.45\columnwidth]{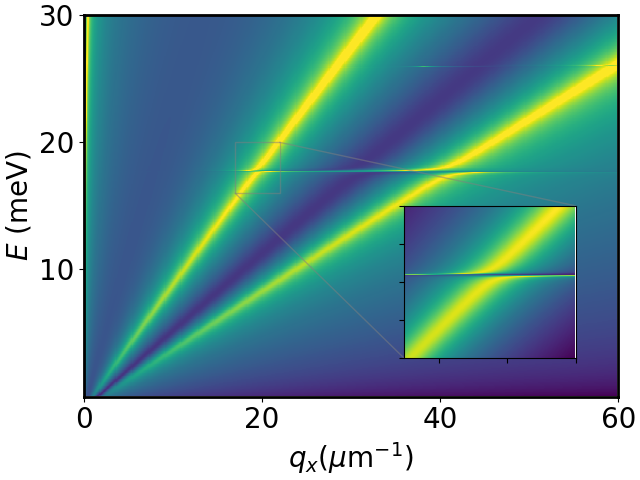} \includegraphics[width=0.45\columnwidth]{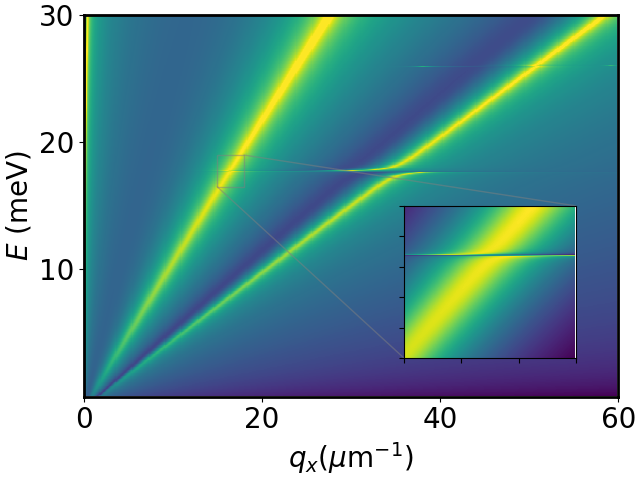}\\
    \includegraphics[width=0.45\columnwidth]{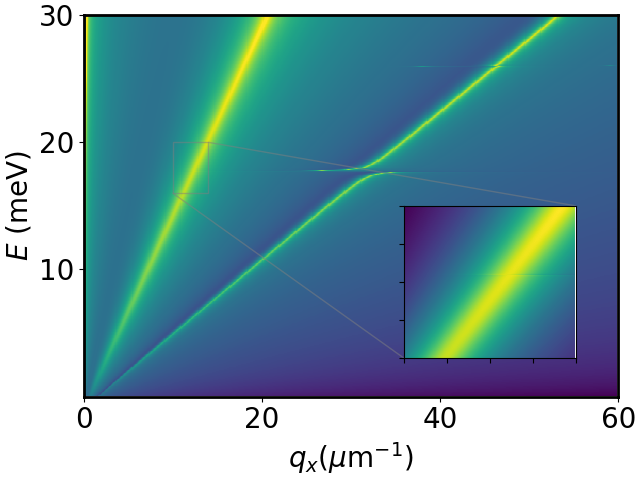}
    \caption{Gr-SC-Gr with different Fermi energies for the two graphene sheets. 
    One of the sheets is kept at $E_F=500$~meV, while the $E_F$ of the other is 
    changed from 50~meV (top left), to 100~meV (top right) to 200~meV (bottom).}
    \label{fig:unbalanced}
\end{figure}

We note that, in this sandwich geometry, only the anti-symmetric graphene plasmon
(the lower energy branch) couples to the superconductor. This is related to
the fact that the electric field at the position of the SC sheet (right 
in the middle of the sandwich) is zero for the symmetric mode. To allow
both branches to couple to the SC we must break the symmetry, either by
placing the SC sheet off center, or by applying different gate voltages
to the two graphene sheets. In figure~\ref{fig:unbalanced} we show the results
for the latter case. It is interesting to note that Fermi energy imbalance between 
the graphene sheets must be relatively large to make the coupling of the high 
energy mode to the SC visible.

\subsection{FeSe - Gr - FeSe}
\begin{figure}
    \centering
    \includegraphics[width=0.4\columnwidth]{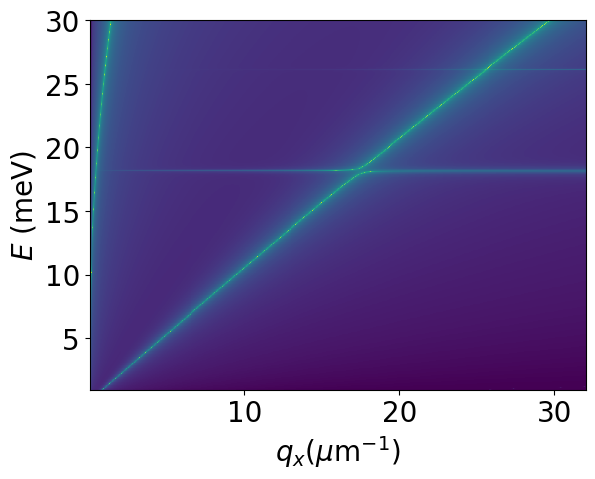} 
    \includegraphics[width=0.425\columnwidth]{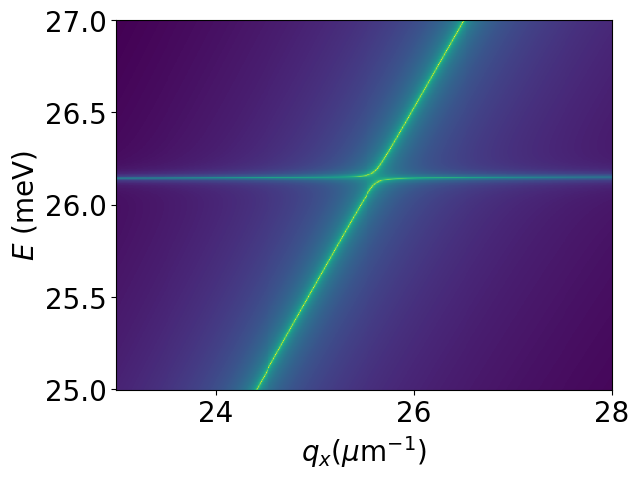} 
    \caption{Imaginary part of the reflection coefficient for a superconducting cavity 
    formed by two sheets of FeSe separated by 12~nm of hBN. Superconducting gap $\Delta=13$~meV, $\hbar\omega_\mathrm{BaSc}(0)=18.2$~meV. The right panel shows a zoom at the region of the crossing  between the Higgs mode and the 2D SC plasmon.}
    \label{fig:FeSe_cavity}
\end{figure}

Superconducting waveguides have been discussed in the literature~\cite{Tsiatmas2012} as
promising building blocks for future plasmonic technologies, due to the
intrinsic low-loss associated with the dynamics of the Cooper-pair condensate.
In the original proposal~\cite{Tsiatmas2012}, the cavity geometry serves the purpose of 
overcoming the problem of weak confinement of the plasmon to the surface of bulk 
superconductors. In the context of 2D superconductors, the same geometry can be
exploited to promote the coupling of the (small dispersion) Bardasis-Schrieffer and Higgs modes to
the anti-symmetric Cooper-pair plasmon, as shown in figure~\ref{fig:FeSe_cavity}. Of course, 
the anti-symmetric mode itself can be used as a resource for plasmonic technologies, as suggested 
in the literature~\cite{Tsiatmas2012}. The features of the anti-symmetric mode can be
tuned in the manufacturing process, by adjusting the distance between the SC monolayers 
and/or changing the insulating spacer material. It may be desirable, however, to have the 
ability to control those features on-the-fly. Incorporating a graphene sheet into the
device may provide such a control. Due to the anti-symmetric nature of the mode, however, 
it couples weakly to objects that are placed close to the middle point between the two FeSe
monolayers. Thus, to obtain control over the dispersion relation of the anti-symmetric mode,
the graphene sheet must be placed much closer to one of the FeSe monolayers than to the other.

\begin{figure}
    \centering
    \includegraphics[width=0.4\columnwidth]{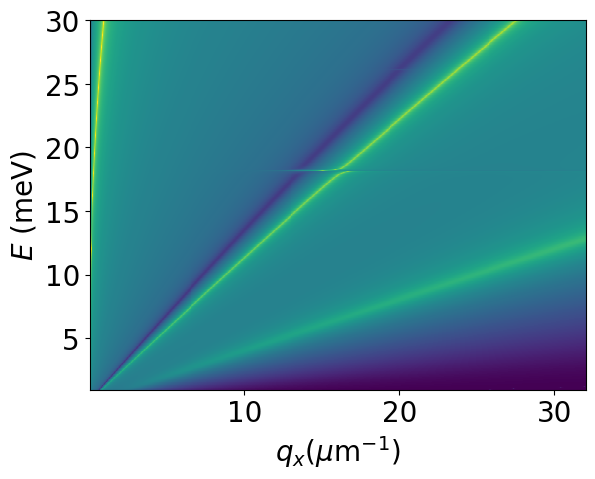}\
    \includegraphics[width=0.4\columnwidth]{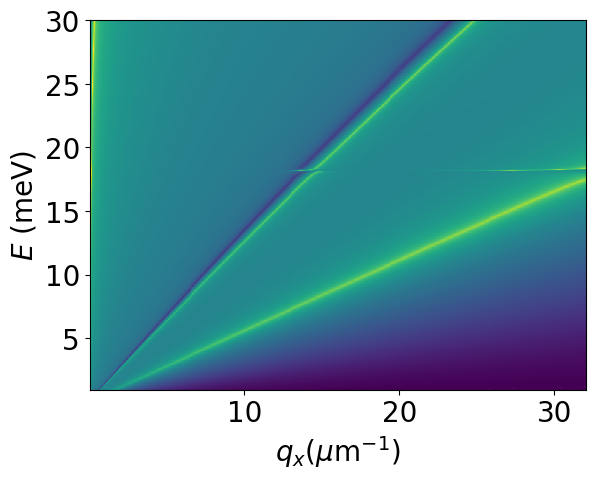}
    \caption{Superconducting FeSe cavity modified by a single graphene sheet placed (asymmetrically) 
    inside the cavity. By changing the doping of graphene it is possible to tune the slope of the anti-symmetric cooper-pair plasmon. Left: $E_F^\mathrm{Gr}=0.1$~eV; right: $E_F^\mathrm{Gr}=0.6$~eV. The distances between the graphene sheet and the left and right FeSe monolayers are 2~nm and 10~nm
    respectively. Superconducting gap $\Delta=13$~meV, $\hbar\omega_\mathrm{BaSc}(0)=18.2$~meV.}
    \label{fig:FeSe_cavity_Gr}
\end{figure}

For the sake of comparison, we start by showing the results for a cavity formed by two 
FeSe monolayers separated by a 12~nm thick slab of hBN. The anti-crossings between the Ba-Sh anti-symmetric plasmon mode is clearly visible. To notice the anti-crossing in the case of 
the Higgs mode it is necessary to zoom in, as shown in fig.~\ref{fig:FeSe_cavity}.
By sandwiching one graphene sheet between the two FeSe monolayers, we add the possibility to tune
certain features of the excitation spectrum of the heterostructure. For instance, by changing
the graphene sheet's doping level it is possible to control the dispersion relation of the 
anti-symmetric plasmon branch associated with the cooper-pairs, as seen in fig.~\ref{fig:FeSe_cavity_Gr}.


\section{Purcell factor}

Direct observation of the hybrid graphene plasmon-superconductor is difficult because
of the mismatch between the wave vector of light and that of the hybrid excitation.
One way to partially circumvent this difficulty is to consider the effects of the hybrid
modes on a nearby quantum emitter~\cite{GoncalvesPeresBook,Koppens2011,Schadler2019,Kurman2018}. 
Its rate of spontaneous emission is modified by the 
presence of the heterostructure, and this modification carries information about the 
reflection coefficients of the heterostructure. This is encoded in the ratio between
the electromagnetic local density of states (LDOS) in the presence of the heterostructure and
the LDOS of free space, given by~\cite{GoncalvesPeresBook}
\begin{equation}
    \frac{\rho(z,\omega)}{\rho_0(\omega)}=1+\frac{1}{2}\int_0^\infty ds\Re\left[\left(\frac{s^3}{s_z}-ss_z\right)e^{2i\frac{\omega}{c}zs_z}r_p(s,\omega),\right],
\end{equation}
where $s_z=\sqrt{1-s^2}$, with $s=q_\parallel c/\omega$ and $z$ is the distance between
the emitter and the topmost hBN layer. This ratio is known as the Purcell factor~\cite{Purcell:1946,Novotny_book,Goncalves_SpringerTheses}, and can be very large when the emitter is placed close to surfaces that support localized electromagnetic modes. 

\begin{figure}
   \includegraphics[width=0.9\textwidth]{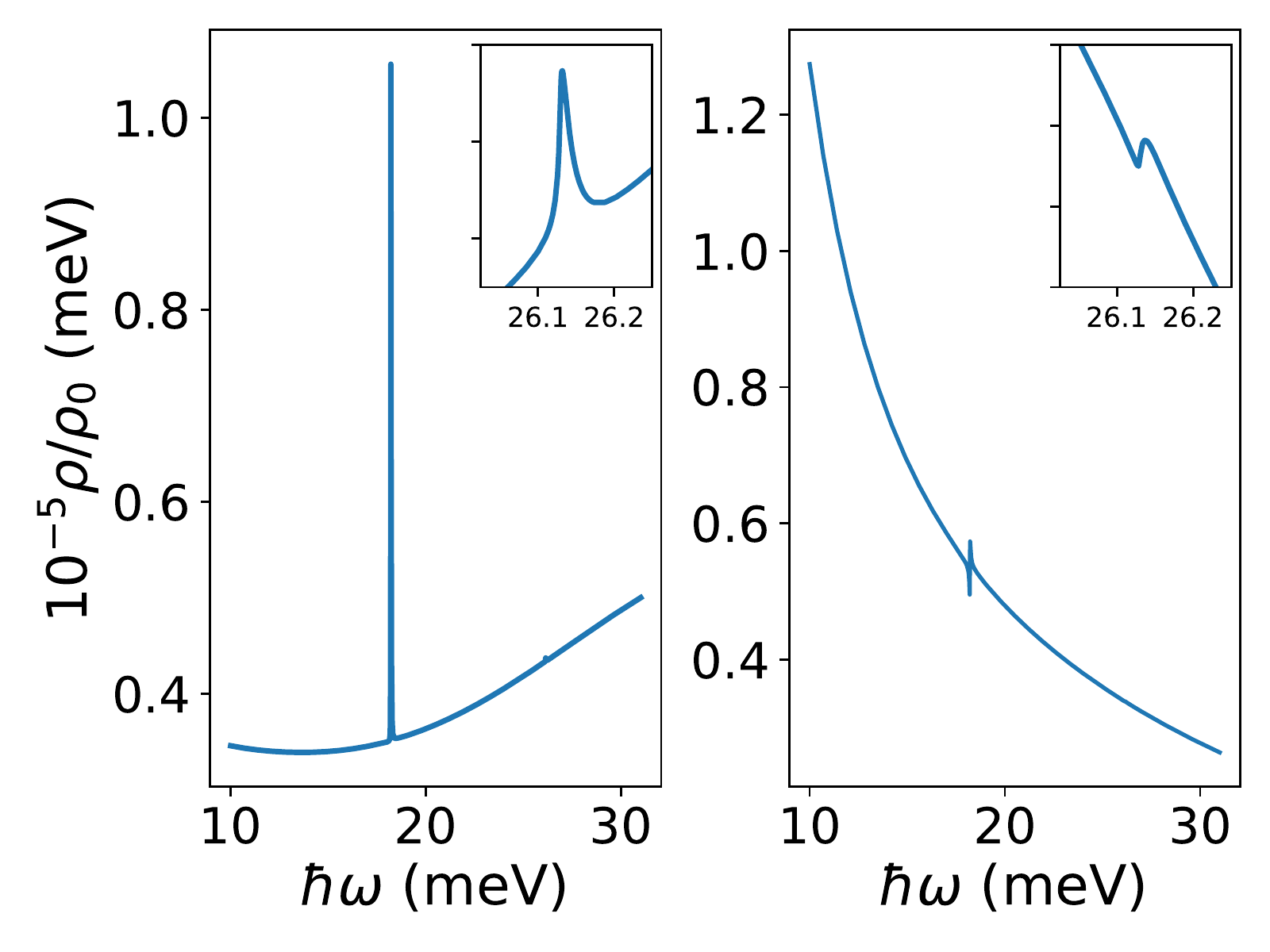}        
    \caption{Purcell factor for the Gr-SC-Gr heterostructure. The left panel shows the limit of
    vanishing graphene doping ($E_F^\mathrm{Gr}=0.1$~meV). For the right panel
    we have chosen $E_F^\mathrm{Gr}=100$~meV. The distance between the graphene sheets and
    the SC is 10~nm. The Purcell factor has been calculated for an emitter placed at a distance of
    50~nm from the surface of the heterostructure.}
    \label{fig:LDOS_0_100}
\end{figure}

In figure~\ref{fig:LDOS_0_100} we show the Purcell factor for a Gr-FeSe-Gr 
heterostructure as a function of the emitter frequency. The emitter has been placed
at a distance $z=50$~nm from the surface, corresponding to an effective wave vector
of $\sim 20\ \mu\mathrm{m}^{-1}$. In the limit of vanishing graphene doping 
($E_F^\mathrm{Gr}=0.1$~meV, left panel) the LDOS displays a clear peak
at the frequency of the Bardasis-Schrieffer mode. It also has a much smaller 
peak at the frequency of the Higgs mode, as expected due to the weakness of the
coupling between the Higgs mode and the electromagnetic field~\cite{sun2020}. In the
right panel we show the Purcell factor for $E_F^\mathrm{Gr}=100$~meV. Both the lineshape
and the size of the features (relative to the background) are strongly influenced by
the charge density in the graphene sheets. These changes reflect the fact that
the hybridization between the SC modes and the graphene plasmon transfers spectral 
weight from a very narrow frequency range (corresponding to the very weakly dispersing 
SC modes) to the hybrid excitations, both of which have significant dispersion.

\begin{figure}
    \centering
    \includegraphics[width=0.9\textwidth]{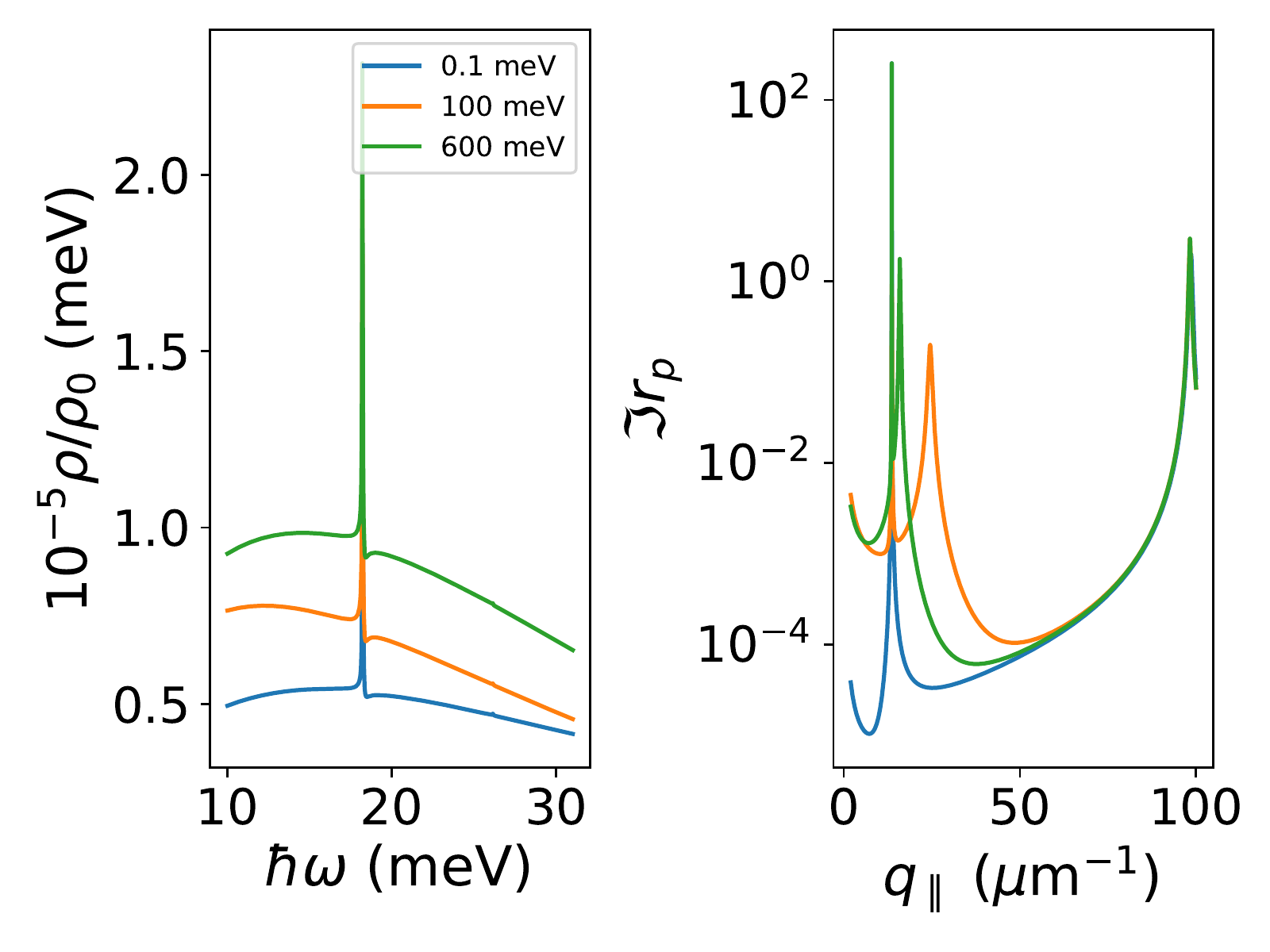}
    \caption{Left panel: Purcell factor for a quantum emitter placed at $z=50$~nm from 
    SC-Gr-SC heterostructure, for three graphene doping levels $E_F^\mathrm{Gr}$. The SC-graphene 
    distance is 10~nm. Right panel: imaginary part of the reflection coefficient as a function
    the wave vector for the same structure and the same doping levels, and $\hbar\omega=18$~meV.}
    \label{fig:LDOS_SCGrSC}
\end{figure}

The Purcell factor for the SC-Gr-SC heterostructure displays a much sharper feature 
when compared to the Gr-SC-Gr structure, as seen in figure~\ref{fig:LDOS_SCGrSC}. 
The Purcell factor in this case is dominated by the very sharp peak in $\Im r_p$
at small wave vector, associated with hybridization between the anti-symmetric 
Cooper pair plasmon and the Bardasis-Schrieffer mode (see the right panel of 
figure~\ref{fig:LDOS_SCGrSC}). The hybridization between the graphene
plasmon and the Bardasis-Schrieffer mode occurs at a larger wave vector
(close to 100 $\mu$m$^{-1}$). Thus, as we will show below, its contribution 
is much attenuated due to the filtering property of the Purcell factor.

\begin{figure}
    \centering
    \includegraphics[width=0.9\textwidth]{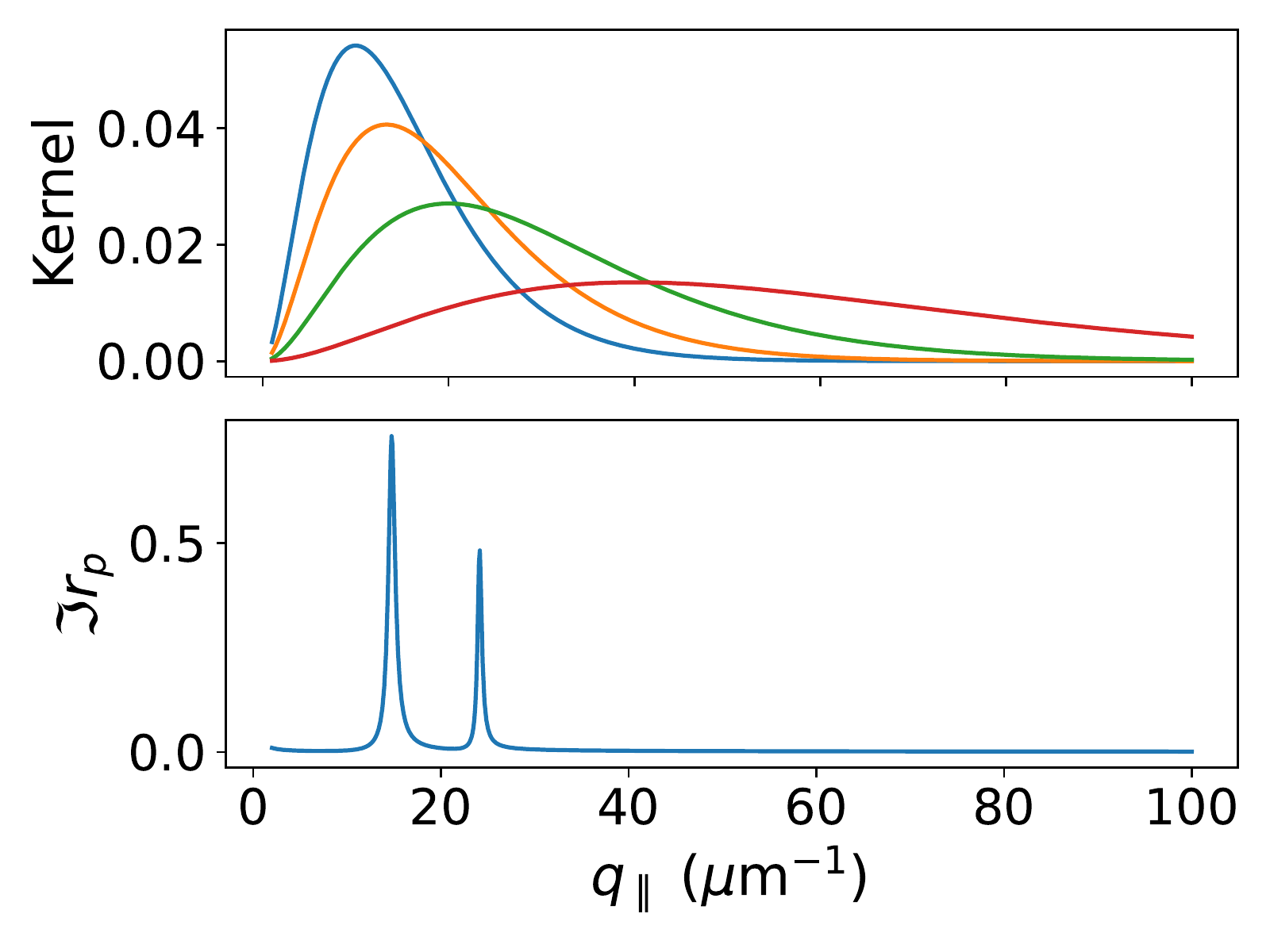}
    \caption{Top panel: kernel (equation~\ref{eq:kernel}) of the integral used to calculate 
    the Purcell factor, for a few values of the distance $z$ between the emitter and the 
    heterostructure. The energy has been fixed at 18~meV. Bottom panel: imaginary part of
    the reflection coefficient for the Gr-SC-Gr heterostructure, for fixed $\hbar\omega=18$~meV; graphene-SC distance is 10~nm, $E_F^\mathrm{Gr}=100$~meV.}
    \label{fig:filter}
\end{figure}

We now comment briefly on the relationship between features
in the near-field reflection coefficient and in the Purcell
factor. Notice that the kernel
\begin{equation}
 \left(\frac{s^3}{s_z}-ss_z\right)e^{2i\frac{\omega}{c}zs_z}
\end{equation}
acts as a filter for the reflection coefficient $r_p(s,\omega)$. In particular,
for $s>1$, $s_z=i\sqrt{s^2-1}$ and
\begin{equation}
    \Re\left[\left(\frac{s^3}{s_z}-ss_z\right)e^{2i\frac{\omega}{c}zs_z}r_p(s,\omega)\right] = \left(\frac{s^2}{\sqrt{s^2-1}}+s\sqrt{s^2-1}\right) e^{-2\frac{\omega}{c} \sqrt{s^2-1} z}\Im r_p(s,\omega).
\end{equation}
This function has a maximum close to $q_\parallel =1/z$, but the width of this peak
depends strongly on both $z$ and $\omega$. To illustrate the filtering property 
of the kernel we fix $\hbar\omega = 18$~meV, which is close to the energy of the 
Bardasis-Schrieffer mode, and plot
\begin{equation}
    K(s,z,\omega)\equiv\frac{1}{N}\left(\frac{s^2}{\sqrt{s^2-1}}+s\sqrt{s^2-1}\right) e^{-2\frac{\omega}{c} \sqrt{s^2-1} z}
    \label{eq:kernel}
\end{equation}
as a function of $q_\parallel$ for a few values of $z$. The normalization
factor $N$ 
\begin{equation}
    N\equiv\int_{1^+}^\infty\left(\frac{s^2}{\sqrt{s^2-1}}+s\sqrt{s^2-1}\right) e^{-2\frac{\omega}{c} \sqrt{s^2-1} z}ds
\end{equation}
is introduced to facilitate the visual comparison between the curves with different values of $z$.
The results are shown in figure~\ref{fig:filter}. It is readily noticeable that the filtering function 
is much broader than the typical features of the reflection coefficient. This constrains the degree
of detail with which features in $\Im r_p$ can be resolved, depending on the wave vector around which
they appear. This is important to keep in mind when choosing the placement of the
LDOS probe relative to the system. 

\section{Concluding remarks}
We have shown that heterostructures of simple geometry, formed by graphene and 
2D superconductors separated by a few nanometers, display clear signatures of 
hybridization between their respective collective modes. The planar cavity geometry 
Gr-SC-Gr promotes a strong enhancement of the hybridization. It also provides tunability, 
either through adjustment of geometric parameters or on-the-fly modification of 
graphene's doping levels. On the other hand, in the SC-Gr-SC cavity the hybridization
is typically weak, but still the graphene doping level can be used to control
the features of the SC collective modes. Our results show that graphene-2D superconductor 
heterostructures are promising platforms for probing the fundamental properties of 2D 
superconductors and for future applications.

\ack
NMRP acknowledges the European Union’s Horizon 2020 under grant agreement  
no. 881603 (Graphene flagship Core3). Additionally, he acknowledges COMPETE 2020, 
PORTUGAL 2020, FEDER and the Portuguese Foundation for Science and Technology (FCT) 
through project POCI-01-0145-FEDER-028114 and UIDB/04650/2020 strategic project.

Both authors acknowledge Frank Koppens, Dimitri Basov, Asger Mortensen, and 
Paulo André Gonçalves for discussions on the topic of the paper.

\section*{References}

\bibliographystyle{iopart-num}
\bibliography{main}

\end{document}